\documentclass[prb,preprint]{revtex4-1}
\usepackage{latexsym}
\usepackage{graphicx}
\newcommand{\figwidth}{2.7 in}

\newcommand{\eexp}{{\rm e}} 
\newcommand{\iima}{{\rm i}} 
\newcommand{\Scal}{{\cal S}} 
\newcommand{\Tcal}{{\cal T}} 
\newcommand{\Ocal}{{\cal O}} 
\newcommand{\Dcal}{{\cal D}} 
\newcommand{\Rcal}{{\cal R}} 
\newcommand{\Tr}{{\rm Tr}}
\newcommand{\Hcal}{{\cal H}}  

\usepackage{float}
\restylefloat{table}

\usepackage{subfigure}
\usepackage{array}
\usepackage{verbatim}
\usepackage{amsmath}
\usepackage{color}
\usepackage{xcolor}
\usepackage{soul}
\usepackage{tabu}
\usepackage{multirow}
\usepackage[normalem]{ulem}
\newcommand{\oldtext}[1]{}
\newcommand{\redtext}[1]{}

\newcommand{\pvec}[1]{\vec{#1}\mkern2mu\vphantom{#1}}

\begin{document}
\graphicspath{{fig/}{./}}
\title{
Density functional theory beyond the Born-Oppenheimer approximation:
 accurate treatment of the ionic zero-point motion
}
\author{Grigory Kolesov$^{(1)}$}
\author{Efthimios Kaxiras$^{(1,2)}$}
\author{Efstratios Manousakis$^{(3,4)}$}
\affiliation{
$^{(1)}$John A. Paulson School of Engineering and Applied Sciences, Harvard University, Cambridge, Massachusetts 02138, USA\\
$^{(2)}$Department of Physics, Harvard University, Cambridge, Massachusetts 02138, USA\\
$^{(3)}$ Department  of  Physics and National High Magnetic Field Laboratory,
  Florida  State  University,  Tallahassee,  Florida  32306-4350,  USA\\
$^{(4)}$Department   of    Physics,   University    of   Athens,
  Panepistimioupolis, Zografos, 157 84 Athens, Greece
}
\date{\today}
\begin{abstract}
We introduce a method to carry out zero-temperature calculations within 
density functional theory (DFT) but without relying on the Born-Oppenheimer (BO)
approximation for the ionic motion. Our approach is based on the finite-temperature 
many-body path-integral formulation of quantum mechanics 
by taking the zero-temperature limit and 
treating the imaginary-time propagation of the electronic variables 
in the context of DFT.  This goes beyond the familiar BO 
approximation and is limited from being 
an exact treatment of both electrons and ions only by the approximations involved in the
DFT component.  We test our method in two simple molecules, H$_2$
and benzene.  We demonstrate that the method produces a difference
from the results of the BO approximation which is significant for many physical
systems, especially those containing light atoms such as hydrogen; in these
cases, we find that the fluctuations of the distance from its equilibrium
position, due to the zero-point-motion, is comparable to the interatomic
distances.
The method is suitable for use with conventional condensed matter approaches
and currently is implemented on top of the periodic pseudopotential code
SIESTA.

\end{abstract}
\maketitle
 \section{Introduction}
The physical properties of solids and molecules
can be determined computationally
by generating many realizations of the system, 
described by its electronic and ionic degrees of freedom,
and sampling the quantities of interest during the numerical simulation. 
A common approach is to separate the electronic and ionic motion,
known as the Born-Oppenheimer (BO) approximation, justified by 
the large mass difference between electrons and ions and the large separation between electronic energy eigenvalues.
Within the BO approximation, the ions may be treated as classical or quantum 
mechanical particles; in either case, an effective interaction potential 
between ions can be obtained by solving the electronic problem for 
each instantaneous ionic configuration, and then using molecular dynamics\cite{BOMD,Marx} 
or Monte Carlo simulations to generate configurations for sampling the system's properties. 

In several situations, a quantum mechanical treatment of the 
ionic degrees of freedom is mandatory.  A case in point is that of 
liquid and solid helium~\cite{ceperley-helium}, $^4$He, 
or helium films on various substrates\cite{Pierce-PRL1,*Pierce-second-layer,*Pierce-solid,*Pierce-corrugations,*Ceperley-Manousakis}.  In these situations,
there are several approaches for capturing the effect of ionic motion by
path-integral Monte Carlo (PIMC), with the electronic degrees of freedom
integrated out through the effective interaction they produce between ions
within the BO approximation.
One can then sample the atomic configurations
using PIMC, as in the original work of Pollock and
Ceperley\cite{Ceperley-1984}.  In the case of $^4$He it has been deemed reasonable to ignore
the electronic degrees of freedom altogether at very low temperature, because
$^4$He is a closed-shell atom in which the first excited atomic state is
several eV above the ground state.
At low temperature, where the average
kinetic energy of the atoms due to their zero-point-motion (ZPM) is of order
$10^{-3}$ eV per atom, they behave as ``elementary'' particles, that is, they
do not exhibit their internal structure as it is extremely unlikely to become
excited through such low-energy collisions.  The effective interatomic
potential in this case can be simply modeled by  a Lennard-Jones type
interaction. Similar empirical-potential and tight-binding path-integral
approaches have been applied in solids
\cite{ramirez1993,kaxiras1994,noya1996,herrero2014path,ramirez2006}. In more general
situations, the disentanglement of the electronic and ionic degrees of freedom
might not be possible\cite{ceperley-hydrogen,graphene_BObreakdown,vidal_LiH,bunker1977,coxon1991,schwenke2001} and accurate approaches have been developed to treat the full electron-ion problem
\cite{tubman2014beyond,ECG_review,h2_wolniewicz_nonadiab, h2_wolniewicz_relat, h2_chen_anderson, yang_diatomic,NEO,NEO-XCHF,h3_pimc}. With these approaches, however, it is presently difficult to go beyond smaller systems. 
A recent
development is a multi-component extension of the density functional theory \cite{kohn-sham} (DFT)
which treats both electronic and nuclei degrees of freedom in the density functional\cite{kreibich_gross,kreibich2008}. The construction of the electron-ion density functional is a difficult problem however and approximations, including the BO approximation, are employed in practice \cite{luders_gross1,luders_gross2}.

A useful and general approach, that has proven quite satisfactory in many applications, 
is to treat only the electrons within density functional theory \cite{kohn-sham} (DFT).
This approach can serve as the basis for
path-integral simulations of ionic motion, where the problem of a quantum
mechanical treatment of ions maps to a classical problem of ring-polymers
\cite{barker1979, chandler_wolynes} interacting by means of the electronic stationary-state energy for the instantaneous atomic configuration of each bead of the
ring polymer 
\cite{Cao-Berne,marx-parinnello,marx-parinnello2,benoit_ice,marx1999nature,morrone_water,stadele_mH,johnson_mH, miller2005quantum, ceriotti, habershon2013ring}.
This formulation is within the BO approximation and ignores the role of
the electronic excitations for a given ring-polymer configuration which
contributes to the path integral over atomic coordinates.

Here, working in the zero-temperature limit,
we introduce an approach that goes beyond the BO approximation and is exact
in the context of the method chosen to solve the electronic problem.
We choose DFT for handling the electronic degrees of freedom, although 
any other approximation with a tractable
time-evolution of the electronic wavefunctions can also be 
implemented in our method.
As far as including the quantum fluctuations of the atomic positions is concerned,  
we use the path-integral formulation.  
In particular, we find the exact eigenstate of the electronic  evolution operator of 
the {\it entire} effective ring-polymer which represents the atomic space-time path in
imaginary time.  This becomes possible because we use the evolution operator
within the DFT formulation that reduces to an effective single-particle-like
evolution, which has to be solved self-consistently. This yields a
self-consistent space-time electronic density, thus incorporating 
``exactly'' within DFT the imaginary-time correlations of the density. As a result, our method
introduces the concept of an  electronic super-wavefunction which is a
space-time-correlated state of the electrons in the entire
pseudo-ring-polymer representing the space-time Feynman path of the atomic
configuration in Euclidean (imaginary) time.
Thus, our choices allow 
us to effectively include the contribution of all virtual electronic excitations.  
Finally, as in other quantum simulation methods, our method employs a
periodic supercell which includes all the atoms for single molecules while 
in the case of crystalline solids it must involve large enough number of primitive unit
cells to limit the role of finite-size effects.  
 
To test the method, we apply it to two model systems, the hydrogen and the
benzene molecules. We find that the size of root-mean-square (rms) 
radius due to the ZPM of the
hydrogen atom is comparable to typical interatomic distances. In this
case, we expect that the evolution of the electronic states and the ionic
motion should be correlated. We also find that the energy difference between our method
and BO approximation-based approaches to this problem is
approximately 5 meV per atom even in the hydrogen molecule that has 
a wide energy gap between occupied and unoccupied electronic states. 
 An energy difference on this scale
can be important in properly describing low-temperature properties and
phases of materials, such as the determination of a charge density wave or
solidification of a system which contains hydrogen or other light atoms.
Furthermore, since life is a subtle phenomenon which is severely affected when
the average energy per atom of the biological system is raised by  $\sim 1$ meV
($\sim 10^{\circ}$K), 5 meV per atom is an energy scale which may have dramatic
effects in living matter.  Since biological systems contain plenty of hydrogen atoms
that participate in important hydrogen-bonded structures, 
their microscopic treatment might benefit from the method presented here.

The paper is organized as follows. In the following Section we present the
method and in Sec.~\ref{implementation} its implementation. In
Sec.~\ref{sec:results} we apply the method to two prototypical smalls systems,
the  H$_2$ and the benzene molecules, and present our conclusions based
on these results in Sec. \ref{sec:conclusions}.
 
\section{Description of the method} \label{method}

The method is described in three steps:
first, the propagation in imaginary-time within the DFT Hamiltonian, next 
the many-body path integral 
form of the partition function within the DFT treatment of the electronic degrees of freedom,
and finally the extraction of
the exact ground-state of the combined ion-electron system within the
DFT scheme.

 \subsection{DFT imaginary-time propagation}

In real-time time-dependent DFT (TDDFT) the time-dependent electronic density
is obtained as the solution to the equation:
 \begin{eqnarray}
  \hat \Hcal^{\rm sp}_{\{\vec R\}}[n(t),\vec r] |\psi_l(t)\rangle &=& \iima \hbar \partial_t
     |\psi_l(t)\rangle,
 \end{eqnarray}
starting from a given initial set of orbitals $|\psi_n(0)\rangle$. Here, for simplicity, the adiabatic approximation is used, that is, the electronic single-particle hamiltonian  $\hat \Hcal^{\rm sp}_{\{\vec R\}}$ is a functional of the instantaneous electronic density  $n(\vec r, t)=\sum_l \psi_l^*(\vec r, t)\psi_l(\vec r, t)$. 
Namely, the single-particle hamiltonian consists of the kinetic energy, the external potential for the
electrons $V_{\rm ext}(\vec r - \vec R_I)$, the Hartree potential $V_H[n,\vec r]$ and the exchange-correlation potential  $V_{xc}[n,\vec r]$ terms:   
\begin{eqnarray}
   \hat \Hcal^{\rm sp}_{\{\vec R\}}[n, \vec r] 
& = & -{{\hbar^2 }\over {2 m_e}} \nabla_{\vec r}^2
   + \sum_{I=1}^{N_{\rm ion}} V_{\rm ext}(\vec r - \vec R_I) \nonumber \\
&+& V_H[n, \vec r]  + V_{xc}[n, \vec r] ,\label{Hsp} \\
V_H[n, \vec r]  &=& e^2 \int \frac{n(\pvec r')}{|\vec r - \pvec r'|} d^3 r'.
\end{eqnarray}
The dependence of the hamiltonian on ionic coordinates, collectively denoted by $\{\vec R\}$, 
is indicated by the subscript.  
The iterative solution to the analytically continued TDDFT equations to imaginary time
 \begin{eqnarray}
   \hat \Hcal^{\rm sp}_{\{\vec R\}}[n(\tau),\vec r] |\psi_l(\tau)\rangle
   = -\partial_{\tau}
     |\psi_l(\tau)\rangle,
 \label{itTDDFT}
 \end{eqnarray}
 where $\tau = \iima t / \hbar $, can be formally written as
 \begin{eqnarray}
    |\psi_l(\tau)\rangle
   = \hat \Tcal \exp \left [ -\int_0^{\tau}  \hat \Hcal_{\{\vec R\}}^{\rm sp}[n(\tau'),\vec r] d\tau'
\right ] 
     |\psi_l(0)\rangle,
 \end{eqnarray}
where $\hat \Tcal$ is the time-ordering operator.
It is straightforward to show that starting from a complete and orthonormal
set of initial states $|\psi_l(0)\rangle$, after infinite imaginary-time $\tau$
(in practice longer than $\hbar /\Delta\epsilon $, where $\Delta \epsilon$ is
the minimum energy-level spacing) the solutions to these equations are the
correct static DFT eigenstates \cite{it_chin,it_mendoza_succi}.  The evolution under imaginary time projects
the lowest energy eigenstate which is not orthogonal to the initial state.
Since we start from a state characterized by definite quantum numbers, which
include the wave-vector $\vec k$ and band index, the minimum energy spacing is
not necessarily zero in the subspace defined by fixing these quantum numbers.

 \subsection{Finite temperature formulation}
 We next wish to calculate the average expectation value of a given observable
 $\hat \Ocal$ as usual
 \begin{eqnarray}
   \langle \langle \hat \Ocal \rangle \rangle  =
           {{\Tr(\hat \rho \hat \Ocal) } \over {\Tr(\hat \rho)}}
\end{eqnarray}
where the trace refers to averaging over all possible ionic configurations
$\{\vec R\}$ and over a complete basis of electronic states. The total
contribution to the statistical density matrix $\hat \rho$ is given as $\hat
\rho =\exp(-\beta \hat \Hcal)$, where $\hat \Hcal$ is the many-body hamiltonian
operator for the ion-electron wavefunction.
The average $\langle \langle \hat \Ocal \rangle \rangle $ 
can be carried out using Feynman paths in imaginary time\cite{feynman,feynman_book},
by writing
\begin{eqnarray}
  \eexp^{-\beta \hat \Hcal} = \eexp^{-\Delta \tau \hat \Hcal} \eexp^{-\Delta \tau \hat \Hcal} ...  \eexp^{-\Delta \tau \hat \Hcal},
\end{eqnarray}
where $K \Delta \tau = \beta$ ($K$ is the number of terms in the above product). 
We can introduce complete sets of states, namely,
\begin{eqnarray}
  \int \prod_{I=1}^{N_{\rm ion}} d\vec R_I^{(j)} | \vec R_I^{(j)} \rangle \langle \vec R_I^{(j)} | \sum_{n_j} | \Psi_{n_j} \rangle \langle \Psi_{n_j} |
  = \hat 1,
  \end{eqnarray}
$K-1$ times, between each $j$th pair of exponentials. We have chosen the electronic states to be the ion-independent states $| \Psi_{n_j} \rangle$, which denote $N_{\rm ele} \times
N_{\rm ele}$ Slater determinants of all possible selections of $N_{\rm ele}$
orbitals from the entire single-particle Hilbert space spanned by a
suitable complete single-particle basis:
 \begin{eqnarray}
     \sum_{n_j} | \Psi_{n_j} \rangle \langle \Psi_{n_j} | &=& \hat 1_{\rm ele},
 \end{eqnarray}
Applying the Trotter expansion for the ionic coordinates and the ionic kinetic energy operator and integrating out intermediate electronic states we get:
\begin{eqnarray}
Z & = & \int \Dcal \vec \Rcal \eexp^{-\Scal_E}
 \sum_{n_1} \langle \Psi_{n_1} | 
 \exp\left(-\Delta\tau \hat \Hcal_{\{\vec R^{(1)}\}} \right)  
 \sum_{n_2} | \Psi_{n_2} \rangle \langle \Psi_{n_2}|
 \exp\left(-\Delta\tau \hat \Hcal_{\{\vec R^{(2)}\}} \right) \nonumber\\
 & & \sum_{n_3} | \Psi_{n_3} \rangle \langle \Psi_{n_3}|
 \ldots
 \exp\left(-\Delta\tau \hat \Hcal_{\{\vec R^{(K)}\}} \right)  
 | \Psi_{n_1} \rangle  
 \nonumber\\
  & = & \int \Dcal \vec \Rcal \eexp^{-\Scal_E}
 \sum_{n_1} \langle \Psi_{n_1} | \prod_{j=1}^K 
 \exp\left(-\Delta\tau \hat \Hcal_{\{\vec R^{(j)}\}} \right)  
 | \Psi_{n_1} \rangle,
 \\
   \langle \langle \hat \Ocal \rangle \rangle  &=& {1 \over Z}
  \int {\cal D}\vec {\cal R}  \hskip 0.1 in \eexp^{-\Scal_E}
	\times \sum_{n_1} \langle \Psi_{n_1} | \hat \Ocal
	   \prod_{j=1}^{K}  \exp\left(-\Delta\tau \hat \Hcal_{\{\vec R^{(j)}\}} \right)
 | \Psi_{n_{1}}  \rangle,\\
{\cal D} \vec {\cal R} &\equiv& \prod_{j=1}^K   \prod_{I=1}^{N_{\rm ion}}  d\vec R_I^{(j)},\\
 \Scal_E  &\equiv&
 \sum_{I=1}^{N_{\rm ion}} \sum_{j=1}^K  {M_I \over {2  \hbar^2 \Delta \tau}}
 \left | \vec R_I^{(j+1)}-\vec R_I^{(j)} \right |^2\label{eqSe}, 
\end{eqnarray}
where $Z$ is the partition function $Z=\Tr[e^{-\beta \hat \Hcal}]$ and $\hat
\Hcal_{\{\vec R^{(j)}\}}$ is the electronic hamiltonian at ionic positions
collectively denoted as $\{\vec R^{(j)}\}$. $\int {\cal D} \vec {\cal R} $
stands for integration over all $K$ time slices.  The path integral is over all
possible ionic paths in imaginary-time which start at $\{ \vec R^{(1)}\}$
and end at $\{\vec R^{(K+1)}\} = \{ \vec R^{(1)}\}$ at imaginary time $\hbar\beta$, 
that is, periodic boundary conditions in imaginary time are imposed. Under
usual circumstances the ionic exchanges have a very small contribution and we
have neglected them for simplicity. They can be
introduced by sampling of the cross-over ionic paths\cite{Ceperley-1984}.

We now use TDDFT to map the many-body to single-particle propagators:
\begin{eqnarray}
\eexp^{-\Delta\tau \hat \Hcal_{\{\vec R^{(j)}\}}} & \rightarrow  &
        \hat T^{(j)}, \nonumber\\ 
\hat T^{(j)} & = & \hat \Tcal \exp \left [ -\int_{(j-1)\Delta\tau}^{j\Delta\tau}  \hat \Hcal_{\{\vec R^{(j)}\}}[n(\tau')] d\tau'
\right ],
\end{eqnarray}
where
\begin{eqnarray}
  \hat \Hcal_{\{\vec R\}}[n]   &=&  \sum_{i=1}^{N_{\rm ele}}
\hat \Hcal^{\rm sp}_{\{\vec R\}}[n, \vec r_i] + \Delta E_{\{\vec R\}}[n] , \\
\Delta E_{\{\vec R\}}[n] &\equiv&    \sum_{I<J} 
\frac{Z^2 e^2}{|\vec R_I - \vec R_J|} - \frac{1}{2} \int  V_H[n, \vec r] 
n(\vec r) d^3 r \nonumber \\
&+& E_{xc}[n] - \int V_{xc}[n, \vec r] 
n(\vec r) d^3 r, \label{deltaE}
\end{eqnarray}
The first term in $\Delta E_{\{\vec R\}}$ is the 
total ion-ion electrostatic repulsion term and the last three
terms are the so-called ``double-counting'' terms, which arise due to auxiliary
nature of the DFT equations \cite{kohn-sham,perdew_chapter,giustino_book}. The final
expression is given by
\begin{eqnarray}
   \langle \langle \hat \Ocal \rangle \rangle  &=& {1 \over Z}
  \int {\cal D}\vec {\cal R}  \hskip 0.1 in \eexp^{-\Scal_E}
	\times \sum_{n_1} \langle \Psi_{n_1} | \hat \Ocal
	   \prod_{j=1}^{K}  \hat T^{(j)}
 | \Psi_{n_{1}}  \rangle,\\
Z &=&\int {\cal D}\vec {\cal R}  \hskip 0.1 in  \eexp^{-\Scal_E}  
	\times \sum_{n_1}
	 \langle \Psi_{n_1}  | \prod_{j=1}^{K} \hat T^{(j)}
	  | \Psi_{n_{1}} \rangle.
  \label{path}
 \end{eqnarray}


\subsection{Exact electronic imaginary-time propagation}

Next we present a method for carrying out an exact propagation of the
electronic state in the many-body path integral. This is practically possible because the
electronic sector is described with DFT.  We implement this is as follows: We
draw a space-time atomic configuration $\vec {\cal R} \equiv \left\{ \vec
R^{(j)} \right\} $ for all $N_{\rm ion}$ ions and at all time slices $j=1,2,...,K$,
that is, for the whole ring polymer.  The space-time atomic configuration is
selected from the Gaussian distribution $\eexp^{-\Scal_E}$.  First, given such
an atomic configuration, we are interested in finding the electronic spectrum,
that is, the eigenstates and eigenvalues of the operator
\begin{eqnarray}
\hat {\bf T}^{(K)}(\vec {\cal R}) & \equiv & \prod_{j=1}^K \hat T^{(j)} 
\label{eqTprod}
\end{eqnarray}
Imagine that we have found the eigenstates 
$|\mathbf \Psi^{(K)}_k (\vec {\cal R})\rangle$ of this operator,
that is,
\begin{eqnarray}
  \hat {\bf T}^{(K)} (\vec {\cal R}) |\mathbf \Psi^{(K)}_k (\vec {\cal R}) \rangle = 
\Lambda_k (\vec {\cal R}) | \mathbf \Psi^{(K)}_k (\vec {\cal R}) \rangle.
\label{eqTLambda}
\end{eqnarray}
where $k$ labels the eigenstate of the whole system. 
Since $\Delta E_{\{\vec R^{(j)}\}}$ does not depend
on the local electron coordinates $\vec r_i$ it can be treated as a 
constant contribution; moreover, while 
this contribution is changing during
the electronic time evolution because the density changes, it does not
affect the electronic wavefunctions.  
Then, we can use these eigenstates, which form a complete set, to calculate
the trace over the electronic degrees of freedom in Eq.~(\ref{path}), instead
of the DFT eigenstates. These eigenstates provide more information about the
electronic states of the entire ``polymer'', that is, the space-time atomic
configuration, as  opposed to using  the eigenstates of one particular
electronic configuration. Eq.~(\ref{path}) takes the following form:
\begin{eqnarray}
   \langle \langle \hat \Ocal \rangle \rangle  &=& {1 \over Z}
  \int {\cal D} \vec {\cal R} \; \eexp^{-\Scal_E} \nonumber \\
&\times& \sum_{k} \Lambda_k
  \langle \mathbf \Psi^{(K)}_{k} (\vec {\cal R} ) | \hat \Ocal
 | \mathbf \Psi^{(K)}_{k} (\vec {\cal R}) \rangle, \\
  Z & = & \int {\cal D}\vec {\cal R}  \; \eexp^{-\Scal_E}  \sum_{k} \Lambda_k (\vec {\cal R}).
  \label{path2}
 \end{eqnarray}
At low temperature only the highest eigenvalue
$\Lambda_{\rm max}(\vec {\cal R} ) $ will contribute, that is, we will have 
$\Lambda_{\rm max}(\vec {\cal R} ) = \exp(-\Scal_{\rm ele}(\vec {\cal R}))$
where we call the quantity $\Scal_{\rm ele}$ the ``electronic action''.
The low temperature limit is equivalent to infinitely long imaginary time, in
which case only the space-time configurations of lowest action contribute.
Thus, we obtain
\begin{eqnarray}
   \langle \langle \hat \Ocal \rangle \rangle  &=& {1 \over Z}
  \int {\cal D} \vec {\cal R} \; \eexp^{-\Scal_E} \Lambda_{\rm max}(\vec {\cal R}) 
\Ocal(\vec {\cal R}), \\
  \Ocal (\vec {\cal R}) &\equiv&
  \langle \mathbf \Psi^{(K)}_0 (\vec {\cal R}) | \hat \Ocal |
  \mathbf \Psi^{(K)}_0 (\vec {\cal R}) \rangle,
  \label{Ome} \\
  Z & = & \int {\cal D}\vec {\cal R}  \; \eexp^{-\Scal_E}  \Lambda_{\rm max}(\vec {\cal R}),
  \label{path3}
\end{eqnarray}
where $|\mathbf \Psi^{(K)}_0 (\vec {\cal R})\rangle$ is the eigenstate which corresponds to $\Lambda_{\rm max}$. It is the lowest-action largest-eigenvalue eigenstate  of the operator $\hat {\bf T}^{(K)}$ and it can be
found by repetitive action of this operator on an initial state until
convergence is achieved; the initial state can be chosen as the DFT ground
state of the atomic configuration at the first imaginary time-slice.  Starting
from any state  $|\mathbf \Psi^{(K)} \rangle$ with non-zero overlap with the
exact $|\mathbf \Psi^{(K)}_0\rangle $, and applying the dimensionless
operator $\hat {\bf T}^{(K)}$ on this state we find 
\begin{eqnarray}
  \lim_{L\to \infty}  \Bigl ({\bf T}^{(K)} (\vec {\cal R} )\Bigr )^L |\mathbf \Psi^{(K)} \rangle = c
  | \mathbf \Psi^{(K)}_0(\vec {\cal R} ) \rangle.
\end{eqnarray}
This is achieved by applying the ``bead'' operator $\hat T^{(j)}$ 
on successive beads and going around the ring-polymer sufficient 
number of times $L$ until convergence. 
We discuss in Sec.~\ref{implementation} how this is done in practice.
After having determined this state, we can calculate the matrix element
of the operator of interest $\hat \Ocal$.
Therefore, we accept the atomic configuration $\vec {\cal R} $
with probability $\Lambda_{\rm max} (\vec {\cal R} )$ 
and we calculate the average of the quantity
$\Ocal(\vec {\cal R} )$ defined by Eq.~(\ref{Ome}), as
\begin{eqnarray}
   \langle \langle \hat \Ocal \rangle  \rangle = {1 \over {N_{\rm conf}}} 
\sum_{\vec {\cal R}} ^{\prime}
      \Ocal (\vec {\cal R} )
   \label{Oave}
\end{eqnarray}
where the prime indicates that the sum is over  
space-time configurations $\vec {\cal R}$  which have been
first selected from the Gaussian distribution $\exp[-\Scal_E]$ and 
were accepted or rejected
according to the probability distribution $\Lambda_{\rm max}(\vec {\cal R})$.

\section{Implementation}
\label{implementation}
\subsection{Imaginary-time dependent DFT}
We implemented the method described above in the 
TDDFT/Ehrenfest
dynamics code TDAP-2.0 presented in Ref.~\onlinecite{kolesov2015real}.  This
code is based on the SIESTA\cite{siesta02} package and employs a numerical
pseudoatomic-orbital basis set.  In such a finite, localized basis set the
imaginary-TDDFT (it-TDDFT) equations become:
\begin{eqnarray}
    \partial_\tau |\psi_l\rangle = 
    -S^{-1}\left(\hat \Hcal^{\rm sp}_{\{\vec R\}}[n,\vec r] 
+ \hat Q\right)|\psi_l\rangle,
\end{eqnarray}
where $|\psi_l\rangle$ is $l^{\rm th}$ KS orbital, $S$ is the overlap matrix
with matrix elements $S_{\mu\nu}=\langle \chi_\mu|\chi_\nu\rangle$
in the basis functions $\chi_\mu$,
and $\hat \Hcal^{\rm sp}_{\{\vec R\}}$ is the KS hamiltonian operator
expressed in this basis, with matrix elements 
$\Hcal^{\rm sp}_{\{\vec R\},\mu\nu} = \langle \chi_\mu |\hat \Hcal^{\rm sp}_{\{\vec R\}} |\chi_\nu \rangle$. 
The matrix $\hat Q$ is the term due to the evolution of the basis 
set in imaginary time, with matrix elements: 
\begin{eqnarray}
    Q^{(j)}_{\mu\nu} = \langle \chi_\mu | \partial_\tau | \chi_\nu\rangle 
    \approx \frac{\vec{R}^{(j)}-\vec{R}^{(j+1)}}{\Delta\tau}\cdot\langle \chi_\mu | 
    \nabla_{\vec{R}^{(j)}} | \chi_\nu\rangle.
\end{eqnarray}
We found that the single-particle propagator $\hat t^{(j)}$ is best approximated through
the self-consistent mid-point exponent\cite{tdprop} computed with the
Pad\'{e} approximant\cite{kolesov2015real}: 
\begin{eqnarray}
  \hat   t^{(j)}\approx \exp \left \{ -\Delta\tau
\left[ S^{-1}\left(\hat \Hcal^{\rm sp}_{\{\vec R^{(j)}\}} 
+ \hat Q^{(j)}\right) \right]_{1/2} \right \},
\label{eq_tdprop}
\end{eqnarray}
where the subscript $1/2$ indicates values taken at the middle of the $j$ time
step and approximated by averaging the initial and final values. This is
equivalent to a second-order Magnus expansion\cite{tdprop}.
After each imaginary time-step the wavefunctions are orthonormalized
with the usual modified Gramm-Schmidt procedure, during which 
the normalization constants are obtained as:
\begin{equation}
\lambda^{(j)} = 
\exp\left ( -\Delta\tau\Delta E_{\{\vec R^{(j)}\}} \right )
\prod_l\sum_{l'}\langle \psi_{l}^{(j)}| \hat t^{(j)} | \psi_{l'}^{(j+1)}\rangle .
\label{smalllambda}
\end{equation}
with $\Delta E_{\{\vec R^{(j)}\}}$ the DFT double-counting and ion-ion repulsion terms, Eq.~(\ref{deltaE}).

The wavefunctions are propagated along the ring for several revolutions, until
self-consistency is achieved.  This is defined as the limit
when the maximum difference
between the density matrix elements belonging to the same bead 
in the current ``lap''
and those in the previous ``lap'' has fallen below a preset cutoff value,
typically set to $\leq10^{-6}$. 
The density matrices of all beads are taken into account.  
Self-consistency is normally reached quickly, typically after
2-3 revolutions. Then $\Lambda_{\rm max}$ can be computed as:
\begin{equation}
    \Lambda_{\rm max}(\vec {\cal R} ) = \prod_j \lambda^{(j)}.
\label{prodlambda}
\end{equation}
We found that the use of a localized basis and the non-linearity
of the TD-DFT hamiltonian can cause large numerical errors in
the propagation if the distance between beads is large, which makes the 
imaginary-time velocity high. To cope with this problem, we introduce a
tolerance distance $d_0$ used in the following sense: 
if the distance between two adjacent beads is larger than $d_0$, the
electronic propagator Eq. (30) is substepped with a reduced time-step.  The
intermediate ionic positions used in the propagator are equally spaced and 
thus the euclidean action term $\mathcal{S}_{E}$ (Eq.(13)), which corresponds to
ionic kinetic energy, is not affected.
The brute-force approach for dealing with the propagator errors is to
decrease $\Delta \tau$ for both the electrons and the ions, however this is
computationally expensive with the current implementations of TDDFT. The
substepping introduces effective sub-beads along each straight-line segment
when it exceeds $d_0$. This is somewhat similar to the use of the
semi-classical action in the
BO path integral methods\cite{ceperley_rev95}, where for the given $\Delta \tau$ it
increases the accuracy in comparison to the primitive action, especially at
higher ``velocities''. Thus, although we introduce substepping as a means of
dealing with the numerical errors of the propagator in Eq.(30), it might also
improve the accuracy of the method for the given $\Delta \tau$ regardless of
these errors.

\begin{figure}
\includegraphics[width=\figwidth]{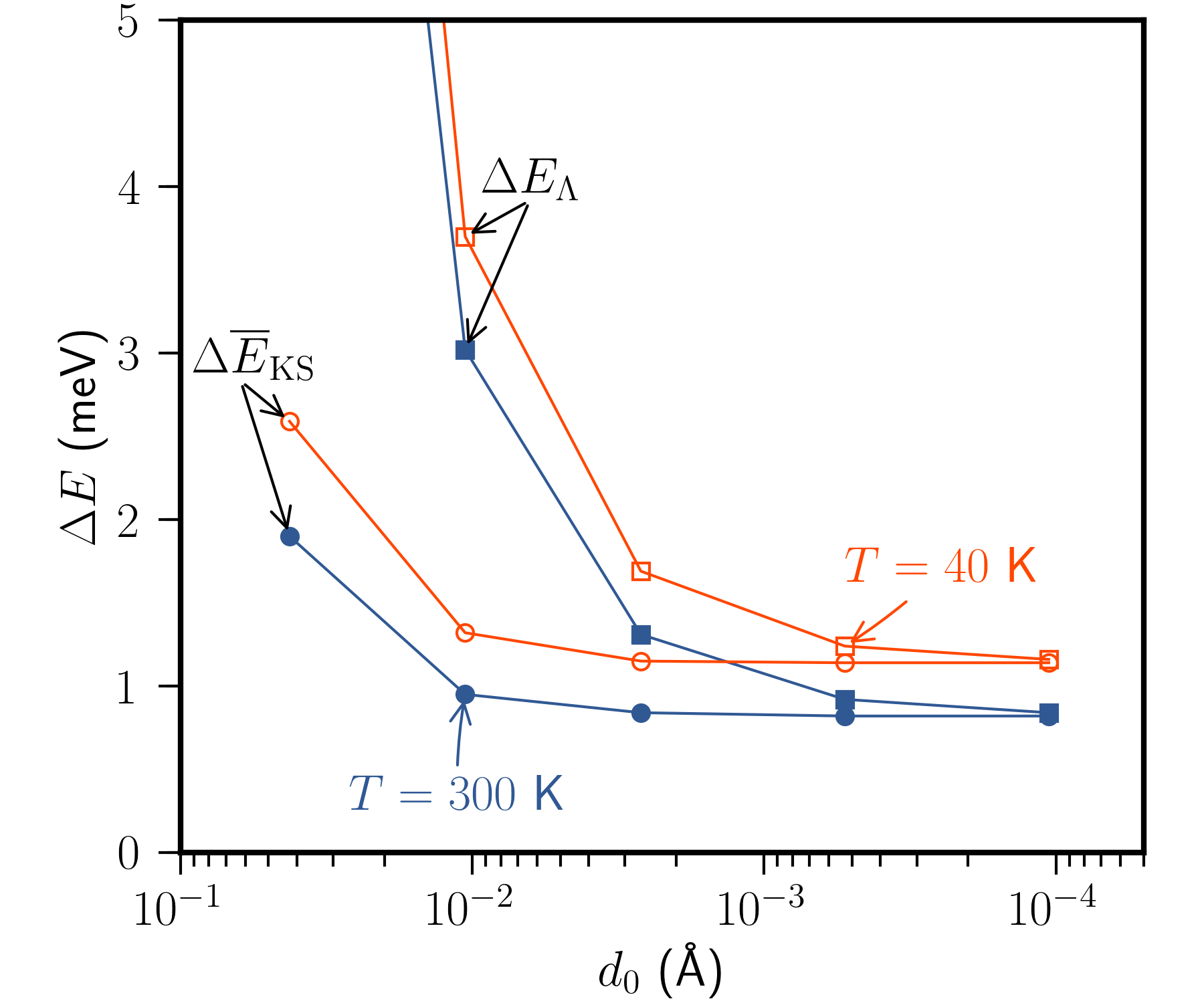}
\protect\caption{Convergence of $\Delta \overline E_{\rm KS}=\overline E_{\rm KS}- \overline E^{\rm BO}_{\rm KS}$ and $\Delta \overline E_{\Lambda}=E_{\Lambda}-\overline E^{\rm BO}_{\rm KS}$ with respect to sub-stepping parameter $d_0$. 
Averaging is performed over the beads (including sub-beads). 
Note, these curves do not represent simulation averages, but rather
energies corresponding to one particular ring configuration randomly drawn from
the MD simulation for the corresponding temperature.
\label{fig:d0conv}
}
\end{figure}

We show the convergence for the single ring
configuration with respect to $d_0$ in Fig.~\ref{fig:d0conv} 
 for the H$_2$ molecule (see also Appendix D, Table~\ref{tab:d0conv} ).  
The ring polymer configurations
for this test were created by running standard adiabatic path-integral
molecular dynamics (PIMD) with the Nos\'{e}-Hoover chain thermostat.
In order to simplify the comparison between the adiabatic and exact
approaches we introduce the energy $E_{\Lambda}$ :
\begin{equation}
E_{\Lambda} = -\frac{\ln{\Lambda_{\rm max}}}{\beta}.
\label{ELambda}
\end{equation}
In the limit of infinitesimally small time slice  $\Delta \tau$, 
$E_{\Lambda} \rightarrow {\overline E}_{\rm KS}$,
where ${\overline E}_{\rm KS} $ is the bead-average of the total electronic
energy 
\begin{equation}
E^{(j)}_{\rm KS}=\sum_{l}
\langle \psi_l^{(j)} |\hat \Hcal^{\rm sp}_{\{\vec R^{(j)}\}}
| \psi_l^{(j)} \rangle +\Delta E_{\{\vec R^{(j)}\}}, \; \; 
{\overline E}_{\rm KS} = \frac{1}{K} \sum_{j=1}^K E^{(j)}_{\rm KS}
\label{Eq:total_E_KS}
\end{equation}  
Fig.~\ref{fig:d0conv} demonstrates that
$E_{\Lambda}$ and ${\overline  E}_{\rm KS} $ indeed converge, and that
${\overline  E}_{\rm KS} $ converges to its final value faster. This convergence
is confirmed in our PIMC runs in section \ref{sec:results} at both $T=300$ K
and $T=40$ K.   

\subsection{Path Integral Monte Carlo}

We implemented the PIMC algorithm that uses staging
coordinates\cite{tuckerman_pimdmc,tuckerman_book}, as reviewed in Appendix A. 
The 
electronic part is treated with it-TDDFT method presented here (denoted as
it-PIMC below) or Born-Oppenheimer DFT (denoted BO-PIMC).
Any average can be obtained by using Eq.s~(\ref{Ome}),(\ref{Oave}).
For the average energy, it is convenient to use the following 
thermodynamic relation: 
\begin{eqnarray}
	E&=&-\frac{\partial}{\partial \beta} \ln{Z} = \nonumber\\
	& =&\lim_{K\rightarrow\infty}
		\left<
			\frac{K D}{2\beta}
		   -\sum_{j=1}^K\left \{\sum_{I=1}^{N_{\rm ion}} \frac{KM_I}{2\hbar^2 \beta^2} 
				\left( \vec{R}_I^{(j)} - \vec{R}_I^{(j+1)}\right)^2 \right \} + 
{\overline E}_{\rm KS} 
		\right>,
	\label{Eave}
\end{eqnarray}
where the average is taken over appropriately distributed configurations
$\{\vec{R}\}$ and $D$ is the dimensionality of the system. The last term inside
angular brackets is obtained with the help of the Hellmann-Feynman theorem (see
Appendix C).
Because for low temperatures the number of time slices required is large
($>200$), some degree of parallel processing is needed even for small systems.
For Born-Oppenheimer  PIMC and PIMD algorithms, parallelism is trivial due
to the independence of the electronic systems at each bead. In the present method 
the electronic systems at different beads are
not independent.  To deal with this problem we first run a long BO
constant-temperature PIMD simulation, in order to generate independent starting
configurations.  By drawing from this set of configurations, a suitable number
of non-adiabatic PIMC simulations can then be started in parallel.

\section{Results\label{sec:results}}

We report results of physical properties of some prototypical systems using our method.
We simulated the H$_2$ molecule at $T=40$ K and $T=300$ K  using the PIMC method,
with $K=381$ and $K=50$ beads, respectively, and using $d_0=0.0026$ \AA~($0.005$ Bohr) in both
cases.  We found that less than 1000 of accepted MC steps are
sufficient for equilibration after sufficiently long thermalization with
staging-coordinate BO-PIMD ($\sim 1.5\cdot 10^5$ steps with 0.05 fs time
steps).
In all cases, we started averaging after 1000 MC
steps.
We used the triple-$\zeta$ plus triple-$\zeta$ polarization shell (TZTP)
basis set and the local density approximation (LDA) with the Ceperley-Alder
(CA) exchange-correlation functional\cite{ceperleyCA} and a standard
pseudopotential from the SIESTA database. Local and semi-local
exchange-correlation functionals such as CA have large self-interaction error
in case of $H_2$ molecule \cite{Pople92}. However we emphasize that the goal of the
simulations here is to compare our method to the standard approaches and not to
the experimental data (for recent  high accuracy experimental measurements of H$_2$ molecule see Refs.
\cite{w_H2,chengH2} and references there in). For this purpose our choices of exchange-correlation
functional and pseudopotential are quite adequate.  We use the same DFT
parameters in all calculations to facilitate this comparison.
The bond length
we obtained after the standard relaxation with the settings and functional
described above is 0.78~\AA. In both BO-PIMC and it-PIMC we obtained about the
same bond length of $\sim$0.81~\AA~for both temperatures
\footnote{
More precisely, the average bond lengths were 
calculated as  0.80995(1)~\AA~and 0.81019(3)~\AA~in it-PIMC at $T=40$ K and $T=300$ K, respectively, and 
 0.80887(1)~\AA~ and 0.80960(1)~\AA~in BO-PIMC calculations at the same values of temperature.
}
(the experimental bond
length for H$_2$ is 0.74~\AA). Although the classical-nuclei bond
length is off by ~0.04 \AA~(as expected for the local
functional\cite{Pople92}), 0.030 \AA~path-integral correction to it is in a reasonable agreement with 0.025 \AA~ correction
obtained in high accuracy calculations \cite{Sims2006,Bubin2003}. 

The results for the zero-point-energy (ZPE) obtained with different  methods are summarized
in Table \ref{tab:H2_ZPE}. 
First we calculate the ZPE using standard Born-Oppenheimer 
harmonic approximation,
with vibrational frequency corresponding to the DFT potential for the H$_2$
molecule.  The harmonic approximation overestimates the ZPE because the high ZPM of the molecule explores the anharmonic region of the potential. 
The energy at $T=40$ K calculated with the Morse potential (with parameters
fitted to match the interatomic potential obtained in our DFT computations)
agrees well with that obtained from BO-PIMC simulations after taking into
account the rotational and thermal motion  using standard rigid rotor
and ideal gas partition functions.
However, these methods underestimate the
ZPE by $\sim 10$ meV in comparison to our exact imaginary-time PIMC (it-PIMC) results.
This correction to BOA agrees well to $14.1$ meV obtained previously
in highly accurate analytic-variational and quantum Monte Carlo calculations of H$_2$ molecule \cite{h2_wolniewicz_nonadiab, h2_wolniewicz_relat, h2_chen_anderson, tubman2014beyond}.  Approximately the same difference is observed between BO-PIMC and
it-PIMC at $T=300$ K, which is not surprising due to the high frequency of
the H$_2$ molecule vibration.
This agreement with the exact calculation is quite good in
  comparison to
$\sim$160 meV  obtained in multi-component DFT
computations\cite{kreibich2008}. Thus our method can also be used to aide the
design of multi-computations density functionals because both methods can be set to
share the same electronic parts of the functional. 
Energy differences between BOA and it-TDDFT in Fig. \ref{fig:d0conv} (see also Table~\ref{tab:d0conv}) 
and in Table \ref{tab:H2_ZPE} differ by one order of magnitude. This is
because in Fig. \ref{fig:d0conv}  the difference is  between two methods
applied to the \emph{same} ring polymer and the \emph{electronic} energy only,
while in Table \ref{tab:H2_ZPE} the differences of the \emph{total} energy
are averaged over a large number of configurations. 
In fact, after decomposing the energy expression of Eq. (\ref{Eave}) into 
nuclear kinetic and
electronic parts we observed that the difference is mostly due to the nuclear
kinetic energy part. 
This can be explained by the following. 
At a given temperature within the BO-PIMC the nuclear kinetic energy and the
electronic energy are partitioned in a certain fraction. The it-PIMC leads to a
repartitioning in which the contribution of the nuclear kinetic energy is
higher as compared to its value in the BO case. This is accomplished in the
it-PIMC procedure by preferring configurations that have, on average, shorter
distances between the beads, which lowers the E$_{\Lambda}$ obtained from
it-TDDFT. This makes the electronic energy close to the BO electronic
ground-state energy.

These results suggest that even for systems like the H$_2$ molecule 
which has a wide gap between occupied and unoccupied energy levels, 
the it-TDDFT correction to the BO approximation is quite 
significant, being roughly 5\% of
the ZPE. We expect these corrections to be larger
for systems with a smaller bandgap and even more so in metallic phases, 
as in the hypothesized high-pressure phase of bulk atomic hydrogen. 

\begin{table}[H]
\renewcommand\thetable{1}
\vskip 0.3 in \centering
\begin{tabular}{|c|c|c|c|c|}
\hline 
$T (K)$ & $E^0_{\rm harm}$ & $E^0_{\rm Morse}$ & $E^0_{\rm BO-PIMC}$ & 
$E^0_{\rm it-PIMC}$ \tabularnewline
\hline 
40 & 279 & 228  & 228.0(2) & 237(1)  \tabularnewline
\hline 
300 & 356 & 312 & 292(1) & 301(1)  \tabularnewline
\hline 
\end{tabular}
\protect\caption{ $E^0 = E^{\rm ZPE} + E^{\rm rot} + E^{\rm COM}$ 
calculated with four methods (all values in meV).
Here $E^{\rm ZPE}$ stands for vibrational zero-point energy, $E^{\rm rot}$ and
$E^{\rm COM}$ are rotational and center-of-mass motion energies at the given
temperature, respectively.
$E^0_{\rm harm}$ is calculated by computing ZPE using the H$_2$ harmonic frequency 
(516.8 meV)  and adding a rotational (rigid rotor) and the center of mass motion 
contributions.  $E^0_{\rm Morse}$ uses the ZPE estimated from a Morse potential
fitted to match the DFT potential energy,
again taking into account the rotational and center-of-mass motion energy.
$E^0_{\rm BO-PIMC}$ and $E^0_{\rm it-PIMC}$ are the energies computed from 
BO-PIMC and
exact it-PIMC simulations, respectively. For MC simulations one standard deviation uncertainty is indicated in parentheses.
\label{tab:H2_ZPE} }
\vskip 0.3 in
\end{table}

\begin{figure*}
\includegraphics[width=15cm]{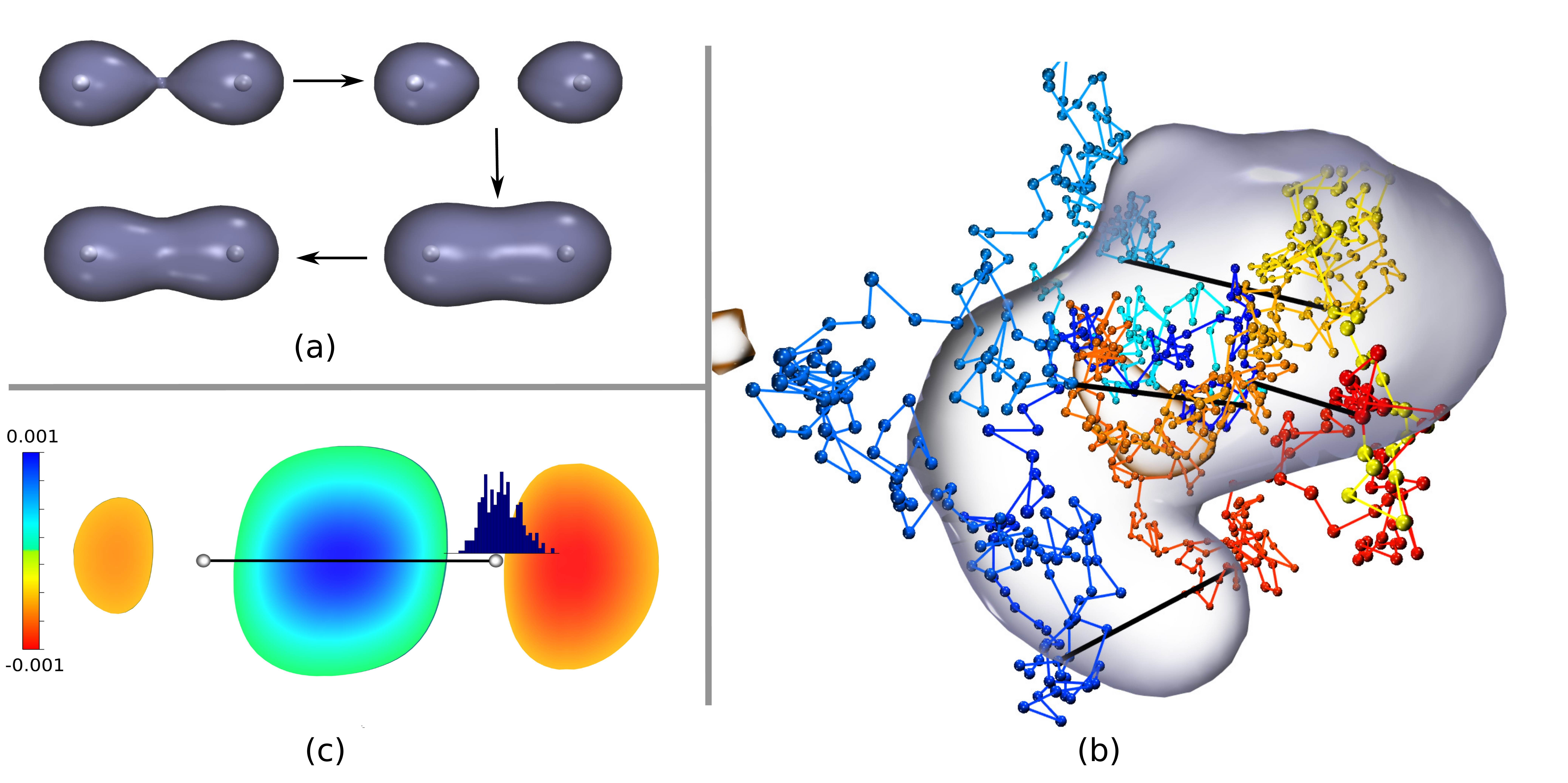}
\protect\caption{
(a) The electron density evolution along the first four beads of the
ring-polymer representing the H$_2$ molecule ($T=40$ K).  
(b) Atomic positions and average density difference between the present method
and the BO approximation for the ring-polymer representing H$_2$. 
Black lines depict the bond between two atoms on the time
slice (shown for intervals of 100 time slices). The bead color corresponds to
time slice, with colors varying from red to yellow for the first atom and
from blue to cyan for the second. 
(c) The average density difference (in units of electrons/\AA$^3$) shown with 
the molecular axis rotated to be in the same direction for all
time-slices by keeping the center of mass of the molecule at the same
position and carrying out the average over all time slices.
Blue-green and red-orange isosurfaces represents the excess of electron
and of hole, respectively.   
The dark-blue distribution above the right atom represents is the
distribution of the distances between two atoms along the ring-polymer in imaginary time.    
}
\label{fig:density}
\end{figure*}

\begin{figure}
    \includegraphics[width=0.5 \textwidth]{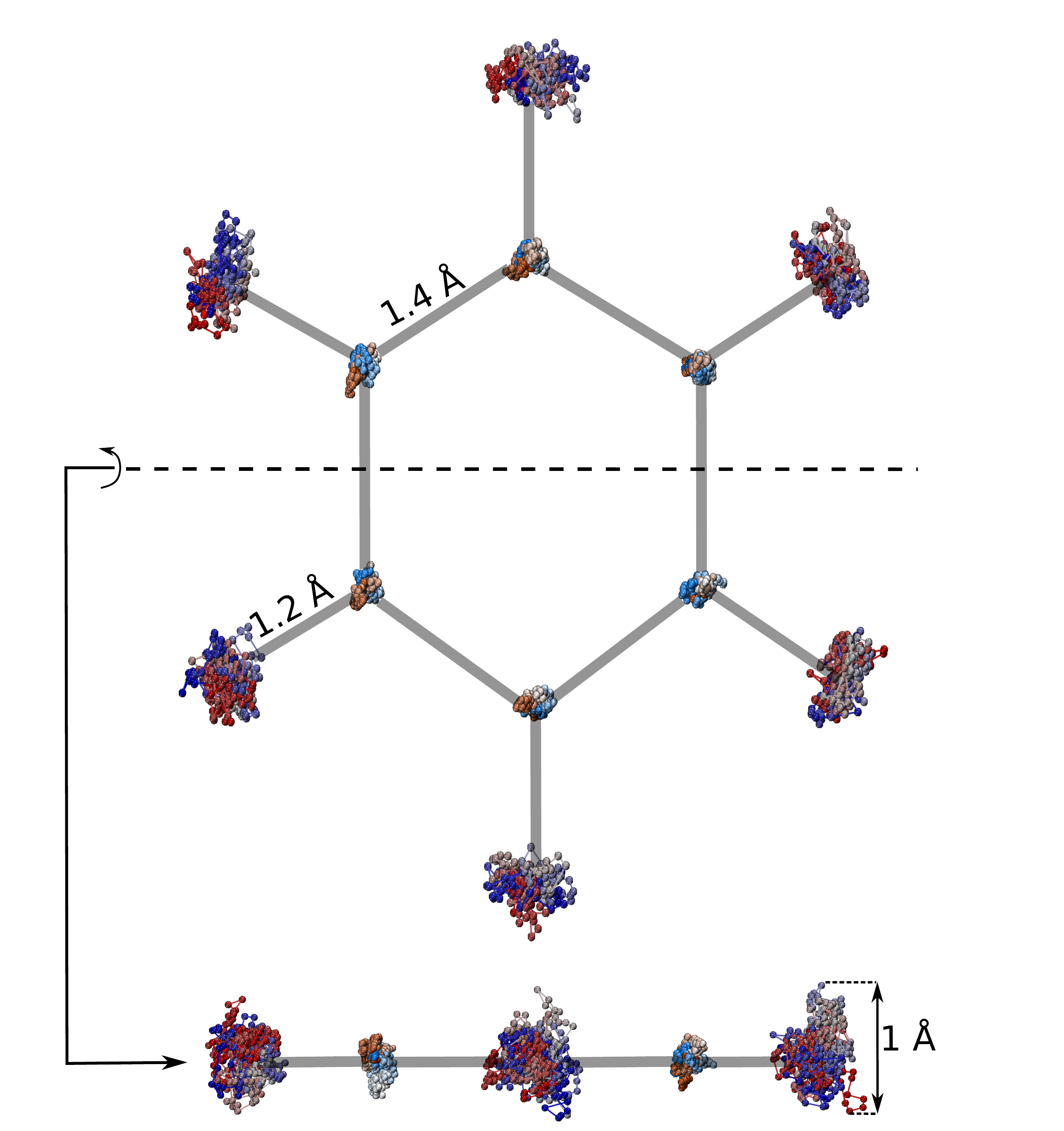}

\protect\caption{The interacting ring-polymers representing the spacetime
  positions of the atoms of the benzene molecule at $T=40$ K, top and side
views. The solid lines connect the equilibrium positions of each atom.  The
same color is used to denote that these positions belong to the same imaginary
time-slice with lighter colors used for carbon atoms.
  \label{fig:benzene}}
\end{figure}

In Fig.~\ref{fig:density}(a) we present the imaginary-time evolution of the electronic density distribution for the first four time-slices for the
H$_2$ molecule. For proper comparison, we rotated
the molecular axis to be in the same direction for all time-slices and kept 
the center of mass of the molecule at the same position.
As the distance between the two hydrogen atoms in the molecule  fluctuates
the electronic density adjusts from one in which the electrons are 
localized at each atom (when the distance between the atoms is relatively large)
to one in which the electrons are shared by the two atoms.
We note that the electronic wavefunctions which 
determine the density are also defined
and evaluated at intermediate times between two successive beads. 
In going from one bead to the next the wavefunction is determined by
evolving the wavefunction which corresponds to the first bead by applying 
the imaginary time evolution operator. 
The final wavefunction is determined for the entire ring-polymer 
simultaneously by applying the evolution operator which corresponds to the
entire ring several times until we obtain convergence.

In Fig.~\ref{fig:density}(b) we give an example of a ring-polymer configuration
of the imaginary-time positions of the two atoms in the H$_2$ molecule.
The size of the rms deviation of each atom from their equilibrium position
is large as compared to the interatomic distance. These atomic
position fluctuations are correlated between the two atoms to a 
significant degree: when one of the atoms moves in a certain 
direction going from one bead to the next, the other atom is more likely to 
move in the same direction by a similar amount. 
In the same plot we also present the averaged difference in the
density distribution obtained with
our method from that obtained by applying the Born-Oppenheimer
approximation, across the space-time configuration in 3D space.  
In our method, we find an enhancement in the density between atoms compared to 
the BO approximation result. 
In Fig.~\ref{fig:density}(c) we present the same difference in the density distribution, after rotating and shifting the molecule so that its center of mass is fixed and the bond is on the $x$-axis. The asymmetry in the electronic
density in Fig.~\ref{fig:density}(c) is due to the fact that
the average is done over a single path in which the center of mass
is moving in imaginary time and that implies each atom moves by different
amount and not necessarily in opposite directions. This has important
implications that we discuss below. 

In Fig.~\ref{fig:benzene} we show
the ring-polymers that represent the space-time configuration of the
carbon and hydrogen atoms in the benzene molecule. As expected, the
positions of the hydrogen atoms have large fluctuation, while the heavier 
carbon atoms have much smaller position fluctuations within the ring-polymer.

\begin{figure*}
     \begin{center}
        \subfigure[]{%
            \label{fig:dipole_h2}
            \includegraphics[width=0.4 \textwidth]{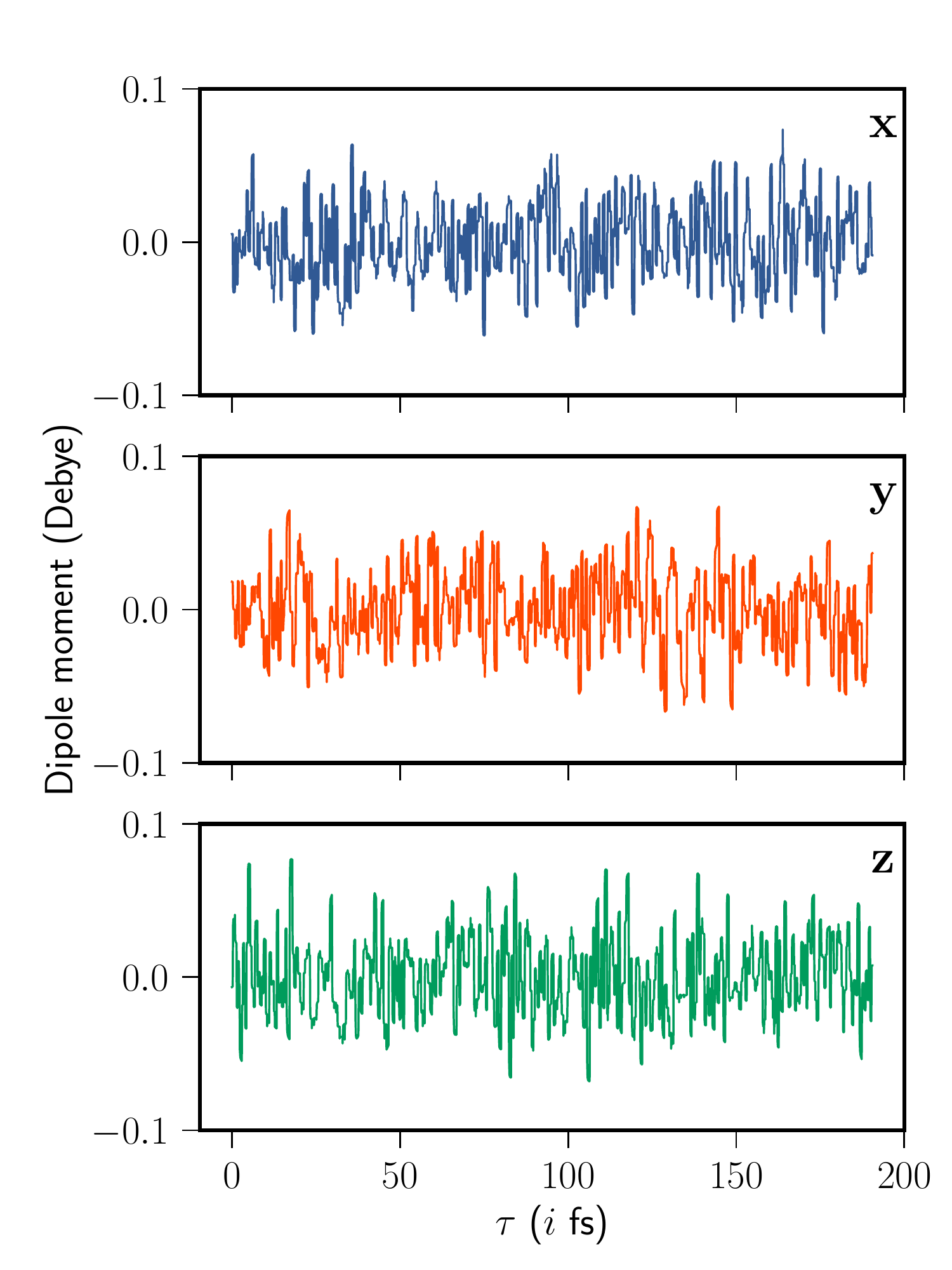}
        }%
        \hskip 0.1 in
        \subfigure[]{%
            \label{fig:dipole_benzene}
            \includegraphics[width=0.4 \textwidth]{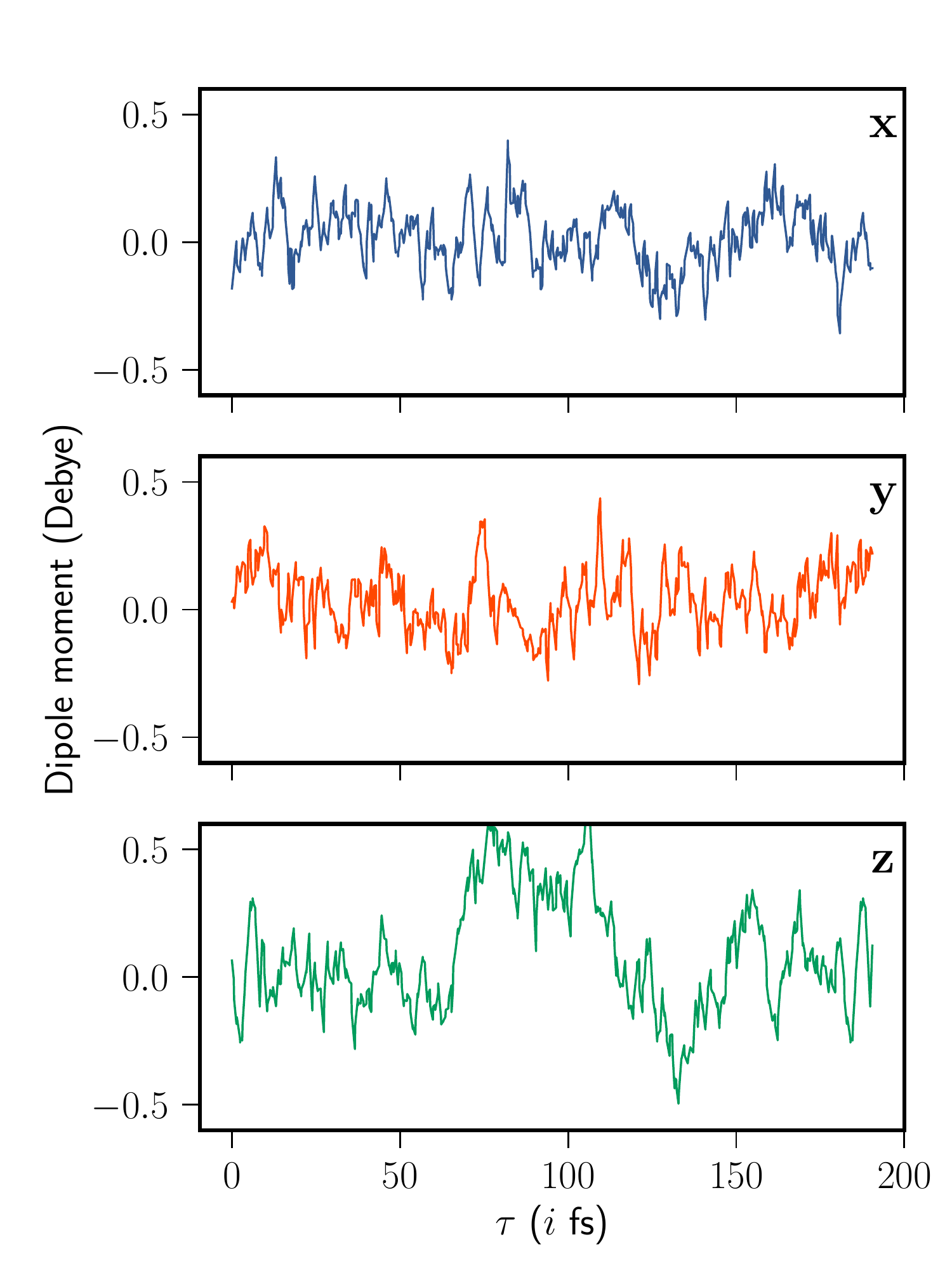}
        }
     \end{center}

\protect\caption{ Imaginary-time evolution of the dipole moment along a single path for (a) the H$_2$ molecule and (b) the benzene molecule. \label{fig:dipole}}
\end{figure*}

Due to the imaginary-time propagation of the electronic degrees of freedom the
mirror symmetry of electronic density is broken. This asymmetry implies a
fluctuation of dipole moment along a single path, which is shown in Fig.~\ref{fig:dipole}
for the H$_2$ and the benzene molecules. The symmetry is restored after summation
over all paths in the case of an isolated molecule. When more than one molecule
is present the fluctuation of the molecular dipole moment can lead to van der
Waals forces between molecules. In our imaginary-time path-integral method 
the presence of van der Waals forces will be manifested by 
increased contribution of the ring-polymer
configurations where the molecular dipole moments are correlated to produce
attraction.  This asymmetry is not present within the BO approximation because
the wavefunctions at each bead correspond to the ground state density
distribution for each bead configuration. In our case, however, we consider the
imaginary-time DFT evolution starting from each bead configuration until we
reach the next, and our wavefunction is an eigenstate of the entire polymer-ring
configuration.

\section{Discussion and conclusions\label{sec:conclusions}}

We have developed an {\it ab initio} method to extend the DFT approach to
include the ionic zero-point motion exactly.  We employ the usual Feynman
path-integral approach where the atomic coordinates form ring-polymers in which
each bead represents the atomic positions at different imaginary-time slices.
The main idea is that we can actually propagate exactly within DFT the
electronic degrees of freedom along each ring-polymer and this allows us to
define a space-time DFT super-wavefunction and electronic density which
characterizes the entire ring-polymer.  This includes imaginary-time
correlations of the electronic state between different beads of the
ring-polymer.  This  exact propagation effectively incorporates the effect of
all virtual electronic states without limiting the description of the
system to the usually adopted BO approximation.

As test cases, we applied our method to the H$_2$ molecule and the benzene
molecule. We find that the difference between our ``exact'' treatment
and the BO approximation is non-negligible when
the system contains light atoms, like hydrogen. This energy difference will
be significant when ionic zero-point motion plays an important role in
determining the prevalent ordered state, like occurrence of  charge density waves. 
Another obvious example of relevance is the
case of highly pressurized hydrogen\cite{ceperley-hydrogen,Silvera}, where 
the zero-point motion is expected to play a significant role, not only
in determining the transition temperature and pressure but, more importantly,
in determining which of the competing phases prevails in the various regimes of
the phase diagram.  We also expect our method to be useful in understanding the
behavior of systems that involve hydrogen bonding and proton exchange, 
which are common in water and in various organic and biological molecules.  
While in general our method requires the entire spectrum of the electronic
system, in systems in which the electronic ground state is separated from the
first excited state by a gap $\Delta$, our method provides information for all
$k_BT \ll \Delta$. Even in the case where in the electronic system the
difference between the ground state and the first excited state is as small as,
say, $\Delta\sim$ 0.3 eV, our approach should be reliable at as high as room
temperature.

Furthermore, the present accurate approach can find application in
several other systems in condensed matter physics where the effects of
correlated motion between electrons and ions is suspected to play an important
role.  Examples include those where there is a Pierls instability where its
formation is assisted by a correlated electron-ion motion, the recently
discovered superconductivity in hydrates\cite{drozdov_H3S}, as well as various polaronic
problems. It is also possible such an approach to find application in
astrophysics, for example, in studies of superconductivity in very cold
brown-dwarf stars (assuming that cold brown-dwarfs exist)  where the electron and ions
should be moving in a correlated fashion.

\section*{Acknowledgments}
This work was supported by  the Army Research Office Multidisciplinary
University Research Initiative (MURI), Award  No.  W911NF-14-0247.   
We used computational resources on Odyssey cluster  (FAS Division of  Science,
Research  Computing  Group  at  Harvard  University) and the Extreme Science
and Engineering Discovery Environment (XSEDE), which is supported by NSF Grant
No. ACI-1053575.

\section*{Appendix} 

\subsection{Path integral Monte Carlo}

In our implementation of PIMC we use the following transformation to staging
coordinates as defined in previous work\cite{Ceperley-1984,tuckerman_book,tuckerman_pimdmc}: 
\begin{eqnarray}
	\vec u^{(j+k)} = \vec R^{(j+W)} - \frac{k\vec R^{(j+k+1)} - \vec R^{(j)}}{k+1},\;\; k=1,2,\ldots,W,
\end{eqnarray}
where $W$ is the segment length (algorithm parameter) and $j$ is a randomly chosen bead. 
The corresponding terms in $\Scal_E$ (Eq. (\ref{eqSe})) transform as:
\begin{eqnarray}
    \sum_{I=1}^{N_{\rm ion}}\sum_{k=0}^W \frac{M_I}{2\hbar^2\Delta \tau} \left | \vec R^{(j+k+1)}_I - \vec R^{(j+k)}_I \right |^2 = \nonumber\\
    \sum_{I=1}^{N_{\rm ion}}\frac{M_I}{2\hbar^2\Delta \tau}\left [ \sum_{k=1}^W \frac{k+1}{k} \left | \vec u^{(j+k)}_I\right |^2  + \frac{1}{W+1} \left |\vec R^{(j+W+1)}_I - \vec R^{(j)}_I \right |^2 \right ] \label{MCgauss}
\end{eqnarray}
This PIMC algorithm  employs two types of moves:
(i) randomly choosing bead $j$ and  drawing new coordinates from
the Gaussian distribution for the transformed coordinates $\vec u$ and (ii) random displacement $\vec{s}$ of the whole ring.
The move is accepted or rejected depending on the ratio of
current (c) and proposed (p)
$\Lambda_{\rm max}$ : if
$q=\Lambda_{\rm max}^p/\Lambda_{\rm max}^c=\eexp^{-\beta(E^p_\Lambda-E^c_\Lambda)}\ge 1$ the move
is accepted, otherwise it is accepted if $r<q$, with $r$ being a uniform
random number in the $[0,1)$ interval. 
$W$ and $s$ are chosen so that the acceptance rate is around 40\%.

\subsection{Path integral molecular dynamics}

For the molecular dynamics sampling method
the staging coordinates are defined by the following relation\cite{tuckerman_book,tuckerman_pimdmc}:
\begin{eqnarray}
    \vec{u}^{(1)} & = & \vec{R}^{(1)} \nonumber\\
    \vec{u}^{(j)} & = & \vec{R}^{(j)} - \frac{(j-1)\vec{R}^{(j+1)}+\vec{R}^{(1)}}{j},\; j=2,...,K.
\end{eqnarray}
Then the partition function, which yields the same averages as the one in Eq.~(\ref{path3}), can be constructed as:
\begin{eqnarray}
    Z  &=&  \eexp^{-\beta \Hcal_{\rm cl}},\\
 \Hcal_{\rm cl} &\equiv& \sum_{j=1}^{K}\sum_{I=1}^{N_{\rm ion}} \left (\frac{[\vec{P}_I^{(j)}]^2}{2{\bar M}_I^{(j)}}
+\frac{1}{2}M_I^{(j)}\omega_K^2 u^{(j)^2} \right )\nonumber \\
		     &+& E_{\Lambda}(\{\vec{R}\}),    \label{Z_MD}
 \\
    M_I^{(1)}&=&0,\ M_I^{(j)}=\frac{j}{j-1}M_I\ (j>1), \\
    {\bar M}^{(1)}_I&=&M_I,\ {\bar M}^{(j)}_I=M_I^{(j)}\ (j>1), \\
    \omega_K&=&\frac{\sqrt{K}}{\beta\hbar}.
\end{eqnarray}
$\Hcal_{\rm cl}$  in Eq. (\ref{Z_MD}) is  a
classical Hamiltonian with  fictitious momenta $\vec{P}_I^{(j)}$ associated
with each bead.  The forces on the beads are derived in the Appendix Section C.  
$Z$ can the be sampled from a standard MD simulation.
To keep the temperature constant we couple every ionic degree of freedom in the
system to Nos\'{e}-Hoover chain or Langevin thermostat.

\subsection{Hellman-Feynman theorem for the ring-polymer \label{axHF} }

In order to derive Eq.(\ref{Eave}) we need to evaluate the derivative of
$\Lambda_{\rm max}$ with respect to $\beta$:
\begin{equation}
\frac{\partial}{\partial\beta} \Lambda_{\rm max} =  
 \frac{\partial}{\partial\beta} \langle \mathbf \Psi^{(K)}_0 | \hat{\mathbf{T}} | \mathbf \Psi^{(K)}_0 \rangle
\label{dbeta1}
\end{equation}
Because $\mathbf \Psi^{(K)}_0$ is the self-consistent eigenvector of $\hat {\mathbf T}$,
\begin{equation}
\frac{\partial}{\partial\beta} \Lambda_{\rm max} =  
  \langle \mathbf \Psi^{(K)}_0 | \frac{\partial}{\partial\beta}\hat{\mathbf{T}} | \mathbf \Psi^{(K)}_0 \rangle
\end{equation}
Then
\begin{eqnarray}
  \frac{\partial}{\partial\beta}\hat{\mathbf{T}} =
      \frac{\partial}{\partial\beta}\prod_j \hat T^{(j)} = 
      \sum_j \hat T^{(1)} \ldots \hat T^{(j-1)}\frac{\partial \hat T^{(j)}}{\partial\beta} \hat T^{(j+1)} \ldots \hat T^{(K)}
\nonumber\\
      = - \frac{1}{K} \sum_j \hat T^{(1)} \ldots \hat T^{(j-1)}\cdot \hat \Hcal_{\{\vec R^{(j)}\}}[n(j\Delta\tau)] \cdot \hat T^{(j)}  \ldots \hat T^{(K)}
\end{eqnarray}
The last equality is derived by writing $\hat T^{(j)}$ directly as
the iterative solution of the it-TDDFT Eq.~(\ref{itTDDFT})
and taking into account the fact 
that $\Lambda_{\rm max}$ is stationary with respect to variation of
 $\mathbf \Psi^{(K)}_0$, therefore terms containing $\int d\vec r\ \delta \Hcal_{\{\vec R^{(j)}\}}/\delta n(\vec r, j\Delta\tau) \cdot \partial n(\vec r,j\Delta\tau) / \partial \beta$ vanish. 
Then,
\begin{eqnarray}
      \langle \mathbf \Psi^{(K)}_0 |  - \frac{1}{K} \sum_j \hat T^{(1)} \ldots \hat T^{(j-1)} \cdot \Hcal_{\{\vec R^{(j)}\}}[n(j\Delta\tau)] \cdot  \hat T^{(j)}  \ldots \hat T_K | \mathbf \Psi^{(K)}_0 \rangle \nonumber\\
      =- \frac{1}{K}\sum_j \lambda^{(1)} \ldots \lambda^{(j-1)}  \langle \Psi^{(j)} |  \Hcal_{\{\vec R^{(j)}\}}[n(j\Delta\tau)] | \Psi^{(j)}\rangle \lambda^{(j)} \ldots \lambda^{(K)} \nonumber \\
      =- \frac{\Lambda_{\rm max}}{K}\sum_j   E^{(j)}_{\rm KS}, 
\label{dbetaN}
\end{eqnarray}
where $\lambda^{(j)}$ is defined by $\lambda^{(j)}|\Psi^{(j)} \rangle =  \hat T^{(j)} | \Psi^{(j+1)} \rangle$, with $\Psi^{(j)}$ being the normalized electronic wavefunction corresponding to $\mathbf \Psi^{(K)}_0$ at the time-slice $j$, see also Eq.s (\ref{smalllambda})-(\ref{prodlambda}). 
Eq.~(\ref{dbetaN}) leads to Eq.~(\ref{Eave}) in the main text.

Similarly, $\vec F_{\rm ele}^{(j)}$, the electronic component of the total force  on the bead $j$,    required for the molecular dynamics sampling algorithm can be derived from
\begin{eqnarray}
\vec F^{(j)}_{\rm ele} = -\nabla_{\vec R^{(j)}} E_\Lambda =\frac{1}{\beta} \nabla_{\vec R^{(j)}} \ln \Lambda_{\rm max}=
\frac{1}{\Lambda_{\rm max}\beta} \nabla_{\vec R^{(j)}} \Lambda_{\rm max}
\label{mdforce1}
\end{eqnarray}
Following the same steps as for Eqs. (\ref{dbeta1}--\ref{dbetaN}) we get
\begin{eqnarray}
 \vec F^{(j)}_{\rm ele} = - \frac{1}{K} \nabla_{\vec R^{(j)}} E^{(j)}_{\rm KS}
\label{mdforceN}
\end{eqnarray}

\subsection{Convergence of $\Delta \overline E_{\rm KS}$ and $\Delta E_{\Lambda}$ 
\label{axtab}
}

For a more quantitative comparison of convergence rates, we 
provide here the values of the quantities 
$\Delta \overline E_{\rm KS}$ and $\Delta E_{\Lambda}$, 
as well as the actual value of the average $\overline E_{\rm KS}^{\rm BO}$, 
for different values of the sub-stepping parameter $d_0$.

\begin{table}[H] 
\centering
\begin{tabular}{|c|c|c|c|c|c|c|}
\hline 
\multirow{2}{*}{$d_0$,Bohr} & \multicolumn{3}{c|}{H$_{2}$, $T=300$ K, $K=36$} & \multicolumn{3}{c|}{H$_{2}$, $T=40$ K, $K=381$} \tabularnewline
\cline{2-7} 
 & $\Delta \overline E_{\rm KS}$ 
 & $\Delta E_{\Lambda}$ 
 & $\overline E_{\rm KS}^{\rm BO} $ 
 & $\Delta \overline E_{\rm KS}$ 
 & $\Delta E_{\Lambda}$ 
 & $\overline E_{\rm KS}^{\rm BO}$ \tabularnewline
\hline 
0.08  & $1.90$ & $12.33$ & -30457.66 & 2.59 & 14.32 & -30519.72 \tabularnewline
\hline 
0.02  & $0.95$ & $3.02$ & -30456.94 & 1.32 & 3.70 & -30520.02 \tabularnewline
\hline 
0.005 & $0.84$ & $1.31$ & -30457.93 & 1.15 & 1.69 & -30520.12 \tabularnewline
\hline 
0.001 & $0.82$ & $0.92$ & -30458.19 & 1.14 & 1.24 & -30520.09 \tabularnewline
\hline 
0.0002 & $0.82$ & $0.84$& -30458.24 & 1.14 & 1.16 & -30520.09  \tabularnewline
\hline 
\end{tabular}

\protect\caption{The data for Fig. \ref{fig:d0conv}. Energies are in meV and $d_0$ is given in units of 
the Bohr radius. Averaging is done over the beads (including sub-beads) for a single 
ring-polymer configuration (see Fig. \ref{fig:density}) . \label{tab:d0conv} }
\end{table}

\bibliography{refs_nourl}

\end{document}